\documentclass[twocolumn,aps,prb]{revtex4} 

\usepackage{graphicx}
\usepackage{amsfonts}
\usepackage{amsmath}
\usepackage{amssymb}
\usepackage{bm}

\newcommand{\Sec}[1]{Sec.\,\ref{#1}}
\newcommand{\App}[1]{Appendix\,\ref{#1}}
\newcommand{\B}{\mbox{\tiny B}}
\newcommand{\M}{\mbox{\tiny M}}

\newcommand{\nl}{\nonumber \\}

\newcommand{\be}{\begin{equation}}
\newcommand{\ee}{\end{equation}}
\newcommand{\bea}{\begin{eqnarray}}
\newcommand{\eea}{\end{eqnarray}}
\newcommand{\bsube}{\begin{subequations}}
\newcommand{\esube}{\end{subequations}}
\newcommand{\Eq}[1]{Eq.\,(\ref{#1})}
\newcommand{\Eqs}[1]{Eqs.\,(\ref{#1})}
\newcommand{\Fig}[1]{Fig.\,\ref{#1}}

\renewcommand{\d}{\mathrm{d}}
\newcommand{\la}{\langle}
\newcommand{\ra}{\rangle}

\begin{document}

\title{Exact dynamics of driven Brownian oscillators}
\author{Rui-Xue Xu} \email{rxxu@ustc.edu.cn}
\author{Bao-Ling Tian}
\author{Jian Xu}
\affiliation{
  Hefei National Laboratory for Physical Sciences at Microscale,
  University of Science and Technology of China, Hefei, Anhui 230026, China}

\author{YiJing Yan}
\affiliation{
  Department of Chemistry, Hong Kong University of
  Science and Technology, Kowloon, Hong Kong}
\date{Submitted to J.\ Chem.\ Phys.\ on July 30, 2008}

\begin{abstract}
 Exact quantum master equation for a
 driven Brownian oscillator system is constructed
 via a Wigner phase-space Gaussian wave packet approach.
 The interplay between external
 field and dissipation
 leads to this system an effective field correction that enhances the polarization.
 This cooperative property is resulted from an effective bath response to the external field
 applied on the system.
 It is important in the low-frequency driving and intermediate bath memory region.
 We demonstrate this non-Markovian effect
 on the linear response and nonlinear dynamics
 and analyze the results together with analytical asymptotic expressions.

\end{abstract}
\maketitle

\section{Introduction}

 Brownian oscillator is the simplest and exactly solvable system in quantum dissipation
theory.\cite{Gra88115,Wei08,Yan05187,Cal83587}
 Physically it serves as an elementary model
 in diversified fields, such as quantum optics, quantum statistics, spectroscopies,
 and chemical dynamics in condensed phases.%
 \cite{Aga692025,Aga71739,Muk95,Kli97,Dit98,Zur03715,Han05026105,Ank05026106,Gra06160,Nit06}
 In this model, the harmonic oscillator ($q$)
 is coupled to a heat bath of harmonic oscillators ($\{x_j\}$) in a bilinear form.
 The total composite material Hamiltonian
 reads\cite{Wei08,Yan05187,Cal83587}
\bsube \label{HM_DBO} \be \label{HM_DBOa}
  H_{\M} = H_0(p,q)+\sum_j\Bigl[\frac{p_j^2}{2m_j}
      +\frac{1}{2} m_j\omega_j^2
         \Bigl(x_j-\frac{c_jq}{m_j\omega_j^2}\Bigr)^2\Bigr] ,
\ee with \be \label{HM_DBOb}
  H_0=\frac{p^2}{2M} +\frac{1}{2}M\Omega^2q^2.
\ee \esube
 The effect of bath is completely
characterized by the interaction bath spectral density
function:\cite{Wei08}
\be \label{Jw0}
 J(\omega\geq 0)
 = \frac{\pi}{2} \sum_j \frac{c^2_j}{m_j\omega_j} \delta(\omega-\omega_j).
\ee

  Note that the Caldeira--Leggett model, \Eq{HM_DBOa},
is intrinsically perturbative, as the
bath is assumed to be harmonic, i.e., $h_{\B}=\sum_j
[p_j^2/(2m_j)+m_j\omega_j^2x_j^2/2]$, and linearly coupled to the
system.  However, it can be treated exactly
without further assumption for an {\it arbitrary} system Hamiltonian,
including anharmonic cases.\cite{Fey63118,Wei08,Tan06082001,Xu07031107}
One can obtain the Feynman--Vernon influence functional
path integral expression\cite{Fey63118,Wei08}
or its equivalent differential
formalism in terms of hierarchical equations
of motion\cite{Tan06082001,Xu07031107}
for the reduced system dynamics.
Great simplification does exist for the Brownian
oscillator where the reduced system of \Eq{HM_DBOb} is also harmonic.%
\cite{Wei08,Hu922843,Kar97153,Haa852462,Hal962012,For01105020}

 The dynamics of Brownian oscillators can be studied in terms of the quantum Langevin equation.
It starts with the Heisenberg equations of motion
for all degrees of freedom, followed by using the formal
solutions to the bath ones.
The resulting quantum Langevin equation reads\cite{Wei08}
\begin{align}
\label{Lang0}
 \dot{\hat p}(t)
 & = - M\Omega^2{\hat q}(t)
      - \int_{t_0}^t \!\!\d\tau\, \gamma(t-\tau) {\hat p}(\tau)
       + F(t-t_0)
 \nl &\quad
      - M  \gamma(t-t_0) {\hat q}(t_0).
\end{align}
Here, $\gamma(t)=M^{-1}\sum_j
(c^2_j/m_j\omega^2_j)\cos(\omega_jt)$ is the friction kernel
function; $F(t) = e^{ih_{\B}t/\hbar}\big(\sum_j
c_jx_j\big)e^{-ih_{\B}t/\hbar}$ is the stochastic force operator,
and its correlation function is related to the
bath spectral density and the friction kernel
functions via the well--established fluctuation--dissipation
theorem.\cite{Gra88115,Wei08,Yan05187}
 Note that if the initial time is set to be $t_0 \rightarrow -\infty$,
the last term in \Eq{Lang0} is reduced to zero, due to the friction
kernel function $\gamma(t\rightarrow\infty)=0$.

 The Brownian oscillator system is more often studied in terms
of the quantum master equation (QME). The key quantity here
 is the reduced density operator $\rho(t)\equiv \mathrm{tr}_{\B} \rho_{\rm T}(t)$,
 i.e., the bath subspace trace of the total density operator.
 The exact QME to Brownian oscillators
 had been constructed via various approaches.
 These include the calculus--on--path--integral
algorithm,\cite{Hu922843,Kar97153}
 the trace--over--bath on total--space Wigner--function
method,\cite{Haa852462,Hal962012,For01105020}
 the stochastic field\cite{Unr891071}
 and the stochastic Schr\"{o}dinger wave function\cite{Str04052115} approaches.
All these cited efforts had not considered the
 effect of time-dependent external fields.

  We have recently constructed an exact QME to
  a driven Brownian oscillator (DBO) system,
 with the assumption of initial thermal equilibrium
before the time-dependent laser field action.\cite{Xu037,Yan05187}
In this work, we will revisit the theory
with an arbitrary initial factorized state.
A single-mode DBO system will be considered for clarity.
 It can be readily extended to multi-mode systems,
 together with the applications to some
 nonlinear spectroscopic problems in condensed
 phases.\cite{Muk95,Yan885160,Cho08011112}

The remainder of the paper is organized as follows.
With the derivation detailed in \App{thWig},
we present the final results
of QME--DBO theory in \Sec{ththeo}.
We discuss its Markovian simplification in \App{app_markov}
and its asymptotic equilibrium property in \App{equi}.
 The new feature of the theory
is an additional effective field correction, resulting
from the interplay between external field interaction and non-Markovian
 dissipation.
 We analyze the effects of the field correction on
the DBO system in \Sec{num},
and finally conclude the paper in \Sec{sum}.

\section{Theory}
\label{ththeo}

\subsection{Model and notations}

The total DBO Hamiltonian
in the presence of an external time--dependent field $\epsilon(t)$
reads [cf.\ \Eq{HM_DBO}]
 \be \label{HTt}
   H_{\rm T}(t) = H_{\M} - q \epsilon(t) =  H(t) + h_{\B} -  \, q \sum_j c_j   x_j.
 \ee
Here, $H(t) = H_0 + H_{\rm ren}  - q \epsilon(t)$ is
the sum of those in the system space,
with the renormalization $H_{\rm ren}=q^2\sum_jc_j^2/(2m_j\omega_j^2)
 =\tfrac{1}{2}M\gamma(0)q^2$. Thus,
 the system Hamiltonian in \Eq{HTt} is
\be\label{Ht}
  H(t) =  \frac{p^2}{2M} + \frac{1}{2}M[\Omega^2+\gamma(0)] q^2  - q \epsilon(t).
\ee
  The effects of bath interaction are described
by the force-force correlation function:
\be \label{Ct}
  C_{\B}(t)\equiv \la  F(t) F(0) \ra_{\B} \equiv \hbar[a(t)+ib(t)].
\ee
Here, $a(t)$ and $b(t)$ denote
the real and imaginary parts of $C_{\B}(t)/\hbar$, respectively;
$\la \, \cdot \, \ra_{\B}
           \equiv {\rm tr}_{\B} (\, \cdot \,  \rho^{\rm eq}_{\B})$
with $\rho^{\rm eq}_{\B}=e^{-\beta h_{\B}}/
           {\rm tr}_{\B}(e^{-\beta h_{\B}})$ being the bath canonical
equilibrium density operator, and $\beta \equiv 1/(k_{\rm B}T)$ the
inverse temperature. The imaginary part is the sine transformation of the
spectral density (with a factor of $-1/\pi$),
and related to
the friction kernel function as $b(t)=\frac{M}{2}\dot\gamma(t)$.
The real part, $a(t)$, depends on temperature
and is related to the imaginary part, $b(t)$,
via the fluctuation--dissipation
theorem.\cite{Gra88115,Wei08,Yan05187}

For later use, we denote the response function
of the Brownian oscillator system as
 \be\label{defchi}
    \chi(t) \equiv i \la[ {\hat q}(t),{\hat q}(0)] \ra_{\M}  /  \hbar \,   .
 \ee
Here, ${\hat q}(t)=e^{iH_{\M}t/\hbar}qe^{-iH_{\M}t/\hbar}$ and
 $\la \, \cdot \, \ra_{\M} \equiv {\rm tr}_{\M} ( \, \cdot \,
 \rho^{\rm eq}_{\M} ) \equiv {\rm tr}_{\M} ( \, \cdot \,
 e^{-\beta H_{\M}} )  /  {\rm tr}_{\M} (e^{-\beta H_{\M}})$,
 defined in the total system and bath composite material space.
 From its definition and \Eq{Lang0} (setting $t_0=0$ there),
we have 
 \be \label{ddotchi_gam}
   \ddot \chi(t) + \Omega^2  \chi(t)
   + \int_0^t\!\d\tau\, \gamma(t-\tau) \dot\chi(\tau) =  0,
 \ee
with $\chi(0)=0$ and $\dot\chi(0)=1/M$.
Let
 \be \label{hatchi}
   \hat\chi(\omega)   \equiv\int_0^\infty\!\!\d t\, e^{i\omega t}\chi(t)   ,  \   \
   \hat\gamma(\omega) \equiv\int_0^\infty\!\!\d t\, e^{i\omega t}\gamma(t) .
 \ee
The frequency--domain resolution of
\Eq{ddotchi_gam} is then
 \be \label{chiwhar}
   \hat\chi(\omega) = \{M[\Omega^2 - \omega^2 - i\omega\hat\gamma(\omega)]\}^{-1}.
 \ee
 Its real and imaginary parts are corresponding to the dispersion and absorption, respectively.
 The response function $\chi(t)$ will be the key quantity
 in the following construction of the QME--DBO theory.

\subsection{Quantum master equation: Exact results}
\label{thQME}

For convenience, we set $t_0=0$ hereafter, and
 assume the total system--bath composite state
 is initially of an uncorrelated form of $\rho_{\rm T}(0)=
 \rho(0)\rho^{\rm eq}_{\B}$. Due to ergodicity the effects of initial
 states vanish completely in the course of time and the system
 relaxes to the thermal equilibrium.
 In \App{thWig} we derive the exact QME--DBO theory
via the modified Yan-Mukamel method
on the Gaussian wave packet (GWP) dynamics.\cite{Yan885160,Xu037,Yan05187}

 The final results are summarized as follows.
\bsube \label{QMEall}
\be\label{finalQME}
  \dot\rho = -\frac{i}{\hbar}[H_{\rm eff}(t),\rho]-{\mathcal R}(t)\rho,
\ee
with
\be \label{Heff}
  H_{\rm eff}(t) = \frac{p^2}{2M}+\frac{1}{2}M\Omega_t^2q^2
   -q[\epsilon(t)+\delta\epsilon(t)],
\ee
and
\be
  {\mathcal R}(t) \rho = \frac{i\Gamma_t}{2\hbar}[q,\{p,\rho\}]
  +\frac{M\zeta'_t}{\hbar}[q,[q,\rho]]-\frac{\zeta_t}{\hbar}[q,[p,\rho]].
 \ee
\esube
The involving time--dependent functions are
\bsube\label{allfuns}
\begin{align} \label{W2}
  &\Omega_t^2 =
     \frac{\ddot\chi^2(t) - \dddot\chi(t)\dot\chi(t)}
          {\dot\chi^2(t)-\ddot\chi(t)\chi(t)}  \, ,
\\
\label{gam}
  &\Gamma_t =
     \frac{\dddot\chi(t)\chi(t) -\ddot\chi(t)\dot\chi(t)}
          {\dot\chi^2(t)-\ddot\chi(t)\chi(t)}  \,  ,
\\
\label{zeta}
  &\zeta'_t =\int_0^t\!\!\d\tau\,\dot\chi(\tau)\xi(\tau;t) ,
\ \ \
  \zeta_t =\int_0^t\!\!\d\tau\,\chi(\tau)\xi(\tau;t) ,
\\
\label{epsil_efft}
  &\delta\epsilon(t)
 = \int_0^t\!\d\tau\,\chi_\epsilon(t-\tau;t)\epsilon(\tau),
\end{align}
\esube
with
\bsube\label{chiep_xi}
\begin{align}   \label{chixi}
   &\xi(\tau;t) \equiv a(\tau)+\!\int_0^t\!\!\d\tau'\,
    a(\tau-\tau')\chi_\epsilon(\tau';t),
\\   \label{chiepsilon}
  &\chi_\epsilon (\tau;t)  \equiv
   M[\ddot\chi(\tau)+ \Gamma_t \dot\chi(\tau)
    +  \Omega_t^2  \chi(\tau)] .
\end{align}
\esube

\subsection{Comments}
\label{th_comm}

 The $H_{\rm eff}(t)$ of
\Eq{Heff}, even in the absence of $\delta\epsilon(t)$,
is neither $H(t)$ [\Eq{Ht}] nor its Caldeira--Leggett
counterpart, $H_0-q\epsilon(t)$ [cf.\ \Eq{HM_DBOb}]. The latter
 is recovered in the Markovian limit; see \App{app_markov}.
 The values of $\chi(0)=\ddot\chi(0)=0$,
$\dot\chi(0)=1/M$, and $\dddot\chi(0)=-[\Omega^2+\gamma(0)]/M$ lead to
$\Omega^2_{t=0}=\Omega^2+\gamma(0)$ and $\Gamma_{t=0}=0$;
i.e., $H_{\rm eff}(t)$ assumes $H(t)$ [\Eq{Ht}] at $t=0$
before dissipation occurs.
When the solvent fluctuation
is much slower than the system response dynamics,
\Eq{ddotchi_gam} is reduced to $\ddot\chi(t)+[\Omega^2+\gamma(0)]\chi(t)=0$;
the relaxation is yet to occur. This implies that  in the slow modulation limit
the bath contributes only to
the static reorganization energy.

 Temperature enters the QME  through
$\zeta'_t$ and $\zeta_t$ via the real part of bath correlation function,
$a(t)\equiv {\rm Re}C_{\B}(t)/\hbar$; cf.\ \Eq{zeta} with \Eq{chixi}.
The asymptotic behaviors of $\zeta'_t$ and $\zeta_t$
are related with the thermal equilibrium variances
$\sigma_{pp}^{\rm eq}$ and $\sigma_{qq}^{\rm eq}$;
see \App{equi}. This guarantees the correct equilibrium property of the QME.

The main feature in the present QME--DBO formalism is the inclusion of
the effective local--field correction $\delta\epsilon(t)$
[\Eqs{epsil_efft} and (\ref{chiepsilon})].
It results from an interplay between driving and non--Markovian
dissipation. In the absence of the external driving field, the above QME
formalism is the same as the previous theory; see, for example,
Refs.\ \onlinecite{Hu922843} and \onlinecite{Kar97153}.

\section{Demonstrations and analysis}
\label{num}

\subsection{The effective field correction}
\label{numa}

\begin{figure}
 \centerline{\includegraphics[width=0.85\columnwidth]{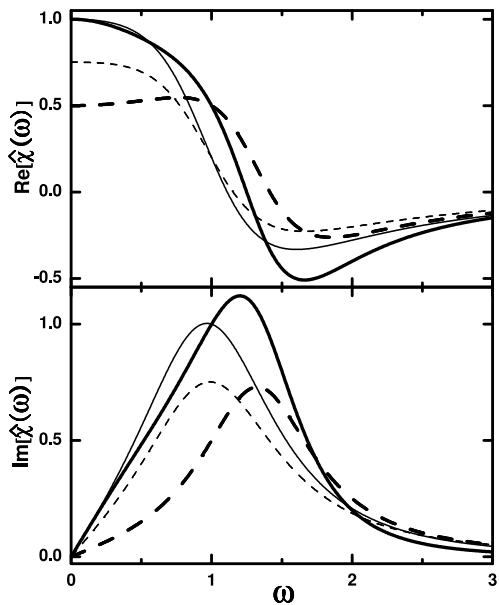}}
 \caption{Real and imaginary parts of the linear response functions
  $\hat\chi(\omega)$ [\Eq{chiwhar}], with the coupling bath strength
  $\Gamma=1$ and two values of spectral width (in unit of $M=\Omega=\hbar=1$):
 $W=2$ (solid) and $5$ (thin-solid).
  The dashed curves are the $\delta\epsilon$--free approximation counterparts.
 }
\label{fig_resp}
\end{figure}

The effective field correction $\delta\epsilon(t)$ is purely of
non-Markovian nature.
(i) It vanishes in the Markovian white noise limit (cf.\ \App{app_markov});
(ii) For a general non-Markovian bath,
if the field correction $\delta\epsilon$ were neglected,
the QME would lead to the system response function $\hat\chi(\omega)$
a Markovian form,
 $[M(\Omega_{t\rightarrow\infty}^2 - \omega^2
 - i\omega\Gamma_{t\rightarrow\infty})]^{-1}$.
 The exact result of \Eq{chiwhar} can only be recovered
with the inclusion of $\delta\epsilon(t)$ in the exact QME--DBO theory;
(iii) $\delta\epsilon(t)$ is most important
in the low frequency driving and intermediate bath memory region;
(iv) The effect of $\delta\epsilon(t)$
is cooperative, leading to the enhancement of polarization.

\begin{figure}
\centerline{\includegraphics[width=0.825\columnwidth]{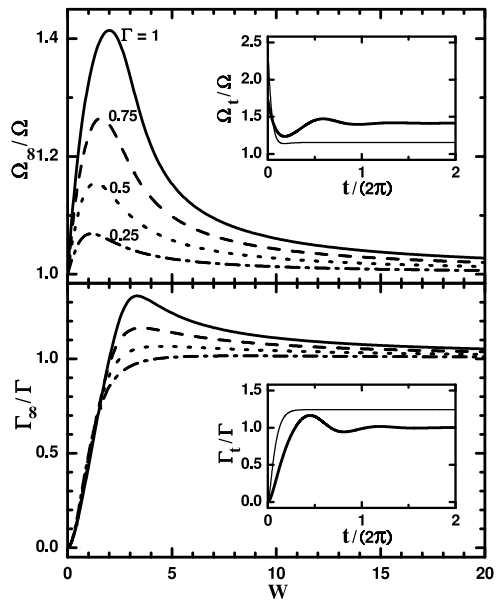}}
\caption{The asymptotic $\Omega_{t\rightarrow\infty}$
  and $\Gamma_{t\rightarrow\infty}$ as functions of $W$
  for $\Gamma=$1 (solid), 0.75 (dash), 0.5 (dot), and 0.25 (dash-dot),
  respectively.
  Shown in the insets are the transient $\Omega_{t}$
  and $\Gamma_{t}$ as functions of time,
  with the coupling bath strength
  $\Gamma=1$ and two values of spectral width,
 $W=2$ (solid) and $5$ (thin-solid).
  All are in unit of $M=\Omega=\hbar=1$.
}
\label{fig_para}
\end{figure}

Note that the asymptotic limit throughout this paper
refers to the $t>t_c$ regime, where $t_c$
denotes the characteristic time of the system response function
having $\chi(t>t_c)\approx 0$.
We adopt hereafter
the internal unit of $M=\Omega=\hbar=k_{\B}=1$,
and the Drude bath model,  $\gamma(t \ge 0) = \Gamma W e^{- W t}$,
for the purpose of numerical demonstrations,
where $\Gamma$ and $W$ denote the bath coupling strength and spectral width, respectively.

Figure 1 compares the system response functions
evaluated with and without $\delta\epsilon(t)$.
The difference between them is considerable
in the intermediate bath memory region,
but vanishes in both
$W\ll\Omega$ (narrow bandwidth) and $W\gg\Omega$ (Markovian) limits (thus not shown).

Before further demonstrations,
let us examine $\Omega_{t}$ and $\Gamma_{t}$,
as these two quantities are
involved in $\delta\epsilon(t)$ [cf.\ \Eq{chiepsilon}],
as well as in the Brownian oscillator part of the QME.
Their initial values are $\Omega_{t=0}=\sqrt{\Omega^2+\gamma(0)}$ and
$\Gamma_{t=0}=0$,
as discussed in \Sec{th_comm}.
Figure 2 depicts the asymptotic values
of $\Omega_{t\rightarrow\infty}$ and $\Gamma_{t\rightarrow\infty}$
varying with the bath spectral width $W$ and coupling strength $\Gamma$,
together with the insets showing $\Omega_{t}$ and $\Gamma_{t}$
as functions of time at $\Gamma=1$, $W=2$ and 5.
The oscillations of $\Omega_{t}$ and $\Gamma_{t}$ vanish at large $W$.
The oscillatory frequency is about that of the system response function.
For the asymptotic behavior,
$\Gamma_{\infty}\rightarrow 0$
when $W\ll\Omega$ and
$\Gamma_{\infty}\rightarrow\Gamma$
when $W\gg\Omega$,
while $\Omega_{\infty} \rightarrow \Omega$ in both limits.

Let us now turn to the local--field correction, $\delta\epsilon(t)$ of \Eq{epsil_efft}.
Note that $\delta\epsilon$
depends linearly on the external driving field $\epsilon(t)$.
We may therefore analyze the effect of the field correction
with a continuous--wave ({\it cw}) drive,
$\epsilon(t)=\epsilon_0\sin(\omega_f t)$,
as an arbitrary field is composed of {\it cw} fields.

\begin{figure}
 \centerline{\includegraphics*[width=0.82\columnwidth]{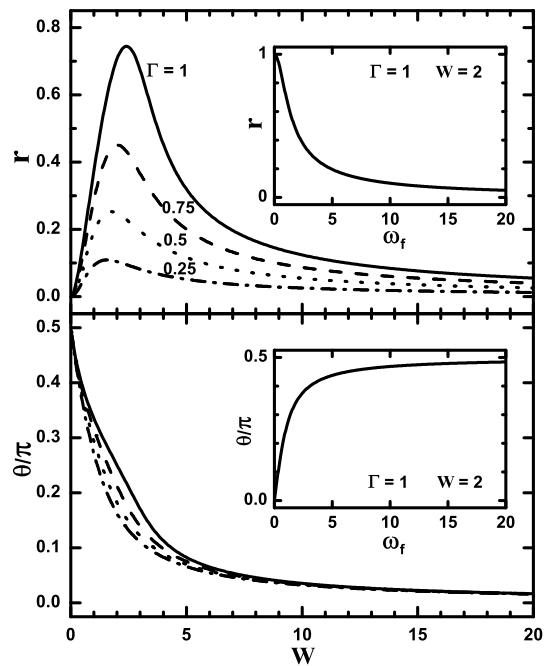}}
 \caption{The amplitude scaling factor $r$ (upper)
 and retarded phase $\theta$ (lower), which characterize
 the asymptotic
 $\delta\epsilon(t)\rightarrow r\epsilon_0\sin(\omega_ft-\theta)$,
 as functions of $W$, with $\omega_f=1$
 and
 $\Gamma=$1 (solid), 0.75 (dash), 0.5 (dot), and 0.25 (dash-dot), respectively.
 Shown in the insets are $r$ and $\theta$ as functions of the {\it cw}
 driving field frequency $\omega_f$,
 evaluated with $\Gamma=1$ and $W=2$.
 All are in unit of $M=\Omega=\hbar=1$.
 }
\label{fig3}
\end{figure}

 Note that the continuous--wave driven Brownian motion had been widely studied
on two-state/double-well systems,\cite{Gri98229,Dit935}
Rydberg atoms in a waveguide,\cite{Blu914521}
as well as dissipative harmonic oscillators,\cite{Zer951533,Koh97300}
with both linear and quadratic dipole moments.
Although the additive term of an effect driving
(that would lead to a form of $\delta\epsilon$)
was mentioned, see for example,
the Appendix B of Ref.\ \onlinecite{Koh97300},
its influence was not studied. And with the
further Markovian approximation made there,
the effective field correction vanishes
and goes unnoticed, as consistent with
the limiting case of the present work;
see  \Fig{fig3} below and the discussion follows.

The asymptotic behavior of $\delta\epsilon(t)$ associating with the
{\it cw}--field can be obtained as [cf.\ \Eqs{epsil_efft} and
(\ref{chiepsilon}) and noting that
$\hat\chi^\ast(\omega)=\hat\chi(-\omega)$] \bsube\label{delf_asy}
\be\label{field_asym}
  \delta\epsilon(t)  \rightarrow  \epsilon_0 {\rm Im}[e^{i\omega_ft}
  \hat\chi_\epsilon^\ast(\omega_f)]
  =r\epsilon_0\sin(\omega_ft-\theta),
\ee
with
\be\label{chiomedef}
 \hat\chi_\epsilon(\omega_f)
 \equiv\int_0^\infty\!\! \d  \tau\, e^{i\omega_f\tau}
 \chi_\epsilon(\tau;t\rightarrow\infty)\equiv re^{i\theta}.
\ee
\esube
In other words, $\hat\chi_\epsilon(\omega_f)$
defines the relative magnitude $r$ and the
retarded phase $\theta$ of the asymptotic
$\delta\epsilon(t)$ with respective to
the original {\it cw} driving field.
Figure 3 demonstrates the
$r$ and $\theta$, for their dependence on the bath spectral width  $W$,
the coupling strength $\Gamma$, and the {\it cw}--field frequency $\omega_f$.

 The relative magnitude $r$ of $\delta\epsilon_{t\rightarrow\infty}$,
as shown in the upper panel of \Fig{fig3},
exhibits turnover behaviors, as function of bath spectral width $W$.
On the other hand, it decreases monotonically as the external {\it cw}--field
frequency $\omega_f$ increases (see the inset of upper panel),
due to the cancellation for the highly oscillatory $\epsilon(t)$.
As the function $r(W)$ is concerned,
both its peak and width increase
rather rapidly as the system--bath coupling strength $\Gamma$ increases.
We may refer the case being of
experimental significance if $r=|\delta\epsilon/\epsilon_0|\ge
0.2$.
With the moderately strong value of $\Gamma=1$,
the field correction would take
a significant effect within
$0.6\Omega \leq W \leq 7\Omega$,
which often covers the experimentally important region.

The retarded phase $\theta$ of $\delta\epsilon_{t\rightarrow\infty}$,
as shown in the lower panel of \Fig{fig3},
decreases from $\pi/2$ at long memory ($W\ll\Omega$) regime,
to 0 in the Markovian ($W\gg\Omega$) limit;
meanwhile it increases from 0 to $\pi/2$ as
the driving field frequency increases; see the inset in the lower panel.
These observations are consistent with the picture that
the retarded phase reflects the relative bath
response time with respect to the inverse frequency of driving.

 We have thus clearly demonstrated that
the  effective field correction $\delta\epsilon(t)$
results from the interplay between the driving
and non--Markovian dissipation.
 Remarkably, the observed retarded phases are all of
$\theta\in[0,\pi/2]$. Thus, the effective field correction is
of {\it cooperative} effect that
enhances the influence of driving.
The physical origin of this enhancement
is due to the bath--induced disperse of $\delta q$ that
leads to an effective larger polarization.
Neglect of $\delta\epsilon$ will underestimate the
influence of driving such as the field--induced
polarizations to be demonstrated below.

\subsection{Driven Brownian oscillator dynamics}
\label{numb}

  Consider now the dynamics of system in the presence
of an external field.
Let us start with the asymptotic behaviors under the {\it cw}--field drive.
The mean values of the coordinate and momentum
can be obtained as [cf.\ \Eq{QtPt}]
\bsube\label{qtptevo}
\begin{align}
q_t&  \longrightarrow  \epsilon_0{\rm Im}[e^{i\omega_f t}{\hat\chi}^\ast(\omega_f)],\\
p_t&  \longrightarrow  \epsilon_0M\omega_f{\rm Re}[e^{i\omega_f t}{\hat\chi}^\ast(\omega_f)].
\end{align}
\esube
The system energy, in terms of $ \la H_0 \ra
=\frac{1}{2M}\la p^2 \ra + \frac{1}{2}M\Omega^2\la q^2 \ra$,
has the asymptotic expression of
\begin{align}\label{energy}
 \la H_0 \ra
    \longrightarrow  &
     \Bigl(\frac{\sigma_{pp}^{\rm eq}}{2M}+\frac{1}{2}M\Omega^2\sigma_{qq}^{\rm eq}\Bigr)+
     \frac{1}{2}M\Omega^2\epsilon_0^2|\hat\chi(\omega_f)|^2
    \nl  &+ \frac{1}{2}M(\omega_f^2-\Omega^2)\epsilon_0^2\{{\rm Re}[e^{i\omega_f t}{\hat\chi}^\ast(\omega_f)]\}^2.
\end{align}
The three terms in the right--hand--side of the above equation
describe, respectively, the thermal average value,
the {\it cw}--field induced static polarization
with the system elongation of $\delta q=|\epsilon_0\hat\chi(\omega_f)|$,
and the residue oscillatory breathing.
In the absence of dissipation, $\hat\chi(\omega)=[M(\Omega^2-\omega^2)]^{-1}$,
both the static polarization and the oscillatory breathing terms diverge at $\omega_f=\Omega$.
This is consistent with the classical energy climbing picture of a {\it cw}--field driven oscillator.
In the presence of dissipation, the oscillatory breathing is asymptotically
quenched at resonance, $\omega_f=\Omega$,
while the static polarization contribution remains finite.
Both the field-induced energy climbing and breathing into
and out of the system are compensated by dissipation.
It is also noticed that the resonance condition is closely related to the choice of reduced system Hamiltonian.
For example, if the reduced system energy is examined in terms of
$\la H(t=0) \ra=\frac{1}{2M}\la p^2 \ra + \frac{1}{2}M\Omega_{t=0}^2\la q^2 \ra$,
similar expression of \Eq{energy} holds, but with the resonance condition being replaced by
$\omega_f=\Omega_{t=0}=\sqrt{\Omega^2+\gamma(0)}$.

\begin{figure}
\centerline{\includegraphics[width=0.815\columnwidth]{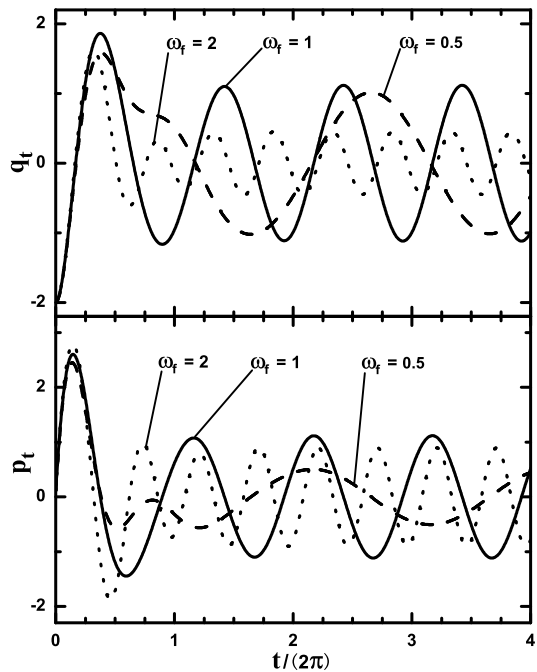}}
 \caption{The evolution of $q_t$ (upper) and $p_t$ (lower)
 under the {\it cw}--field $\epsilon(t)=\epsilon_0\sin(\omega_f t)$,
 with $\epsilon_0=1$ and
 $\omega_f=1$ (solid), 0.5 (dash), and 2 (dot), respectively.
 The bath parameters are $\Gamma=1$ and $W=2$.
 All are in unit of $M=\Omega=\hbar=1$.
 }
\label{fig4}
\end{figure}

\begin{figure}
 \centerline{\includegraphics[width=0.85\columnwidth]{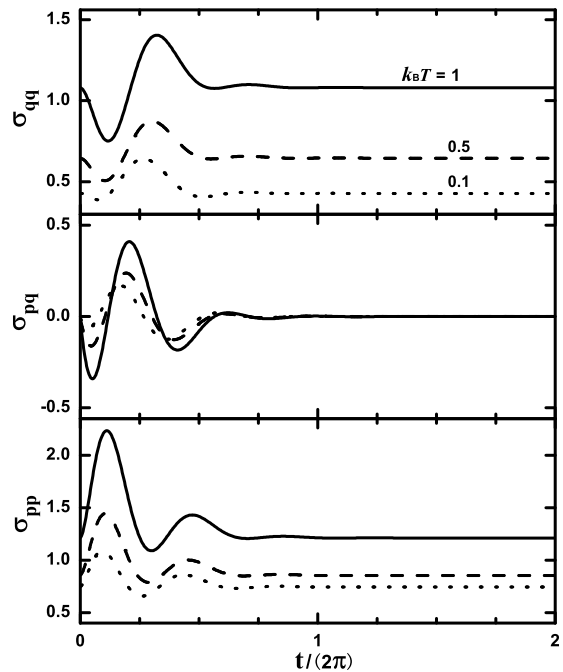}}
 \caption{The evolution of the phase--space variances,
 $\sigma_{qq}(t)$, $\sigma_{pq}(t)$,
 and $\sigma_{pp}(t)$,
 at temperature $T=1$ (solid), 0.5 (dash), and 0.1 (dot).
 The bath parameters are $\Gamma=1$ and $W=2$.
 All are in unit of $M=\Omega=\hbar=k_{\B}=1$.
 }
\label{fig5}
\end{figure}

\begin{figure}
 \centerline{\includegraphics[width=0.85\columnwidth]{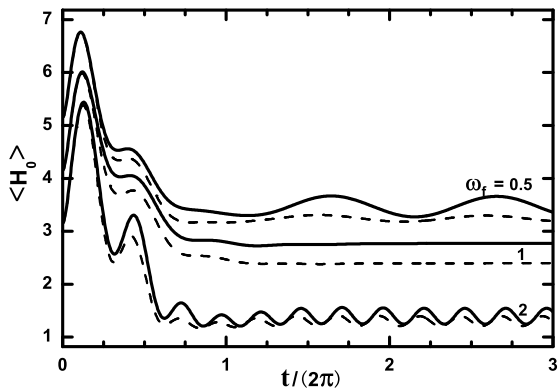}}
 \caption{The evolution of the mean energy $\la H_0\ra$,
 driven by the {\it cw}--field with $\epsilon_0=1$
 and $\omega_f=$0.5 (shifted by $+\,2$),
                  1 (shifted by $+\,1$), and 2 (no shift).
 The bath parameters are $\Gamma=1$ and $W=2$ at temperature $T=1$.
 The dashed curves are the $\delta\epsilon$--free approximation counterparts.
  All are in unit of $M=\Omega=\hbar=k_{\B}=1$.
 }
\label{fig6}
\end{figure}

\begin{figure}
 \centerline{\includegraphics[width=0.85\columnwidth]{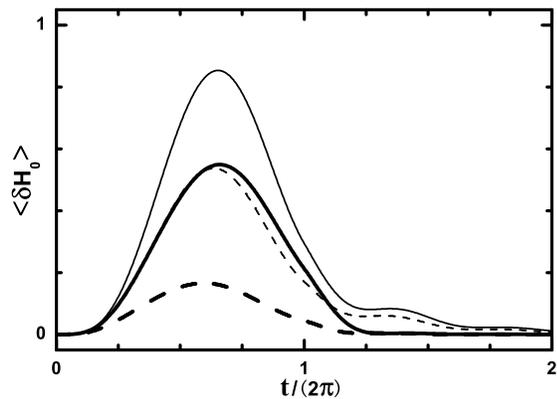}}
 \caption{The evolution of $\la \delta H_0 \ra
  = \la H_0 \ra - \la H_0 \ra_{\rm eq}$,
  driven by a half cycle of the {\it cw}--field with $\epsilon_0=1$
  and $\omega_f=$0.5 on an initial thermal equilibrium state
  at temperature $T=1$.
  The bath parameters are $W=2$, and $\Gamma=1$ (solid) and $0.5$ (thin-solid), respectively.
  The dashed curves are the $\delta\epsilon$--free approximation counterparts.
 All are in unit of $M=\Omega=\hbar=k_{\B}=1$.
 }
\label{fig7}
\end{figure}

 Demonstrated below will be the time evolution of DBO systems,
with both a {\it cw}--field (cf.\ Figs.\,\ref{fig4}--\ref{fig6})
and a half--cycle of the field (cf.\ \Fig{fig7}).
The results will be reported
in terms of the first-- and second--order cumulants,
$\{q_{t},p_{t}\}$ and $\{\sigma_{qq}(t),\sigma_{pq}(t),\sigma_{pp}(t)\}$,
that fully characterize the GWP of the DBO system (cf.\ \App{thWig}),
and also in terms of  $\la H_0 \ra$ or polarization that is experimentally
observable.
Note that $\{q_{t},p_{t}\}$ do not depend on temperature,
while $\{\sigma_{qq}(t),\sigma_{pq}(t),\sigma_{pp}(t)\}$
do not depend on the driving fields;
and their equations of motion, \Eq{dotQtPt} and \Eq{dotW},
are independent.

 For the DBO system in the presence of a {\it cw}--field
$\epsilon(t>0)=\epsilon_0\sin(\omega_ft)$,
we set the initial GWP to have the first--order
cumulants of $\{q_{t=0},p_{t=0}\}=\{-2,0\}$
and the second--order cumulants the same as their thermal
equilibrium values. This mimics the
scenario of ultra-fast UV/visible laser pulse excitation of an
electronic ground state Gaussian wave packet onto a linearly
displaced excited surface, immediately followed by  the {\it cw} infrared field.
The wave packet dynamics on the excited surface under the
influence of the {\it cw}--field is then studied.

Figures 4 depicts the evolution of
$q_t$ and $p_t$. Their asymptotic behaviors, as described by \Eq{qtptevo},
oscillate with the {\it cw}--field frequency $\omega_f$.

Figures  5 demonstrates the evolution of
$\sigma_{qq}$, $\sigma_{pq}$, and $\sigma_{pp}$.
Although they were initially of the equilibrium values
and they are independent of the external field afterward,
the second--order cumulants do oscillate and take time back to equilibrium.
This is because there was no excited state wave packet
initially and the dissipation considered here
only starts when there is a real occupation upon excitation.
Therefore the variances of the pumped GWP, as depicted in \Fig{fig5},
do experience {\it dynamics solvation}.

Figure 6 shows the evolution of the system energy, $\la H_0 \ra
=\frac{1}{2M}\la p^2 \ra + \frac{1}{2}M\Omega^2\la q^2 \ra$.
The asymptotic behavior had been
described in \Eq{energy} and analyzed thereafter.
In the non-resonance condition, $\omega_f\neq\Omega$,
the energy flows in and out with the {\it cw}--field frequency $\omega_f$.
This energy breathing is quenched at resonance, $\omega_f=\Omega$.
Shown in the dashed curves are the $\delta\epsilon$--free counterparts
for comparison.
Their difference
is proportional to the field strength,
and also depends on the difference between the exact $\hat\chi(\omega_f)$
and its approximation; cf.\ \Eq{energy} and \Fig{fig_resp}.

 Consider now the DBO dynamics driven by a half-cycle of the sine-function field.
The system is assumed to be initially at the thermal equilibrium.
Figure \ref{fig7} reports the resulting dynamics
in terms of
the polarization/magnetization related
quantity, $\la \delta H_0 \ra = \la H_0 \ra - \la H_0 \ra_{\rm eq}$.
Again it shows that the effect of $\delta\epsilon$
is cooperative,
and the neglect of this correlated driving and non--Markovian
dissipation contribution will
significantly underestimate the field--induced polarization.

\section{Summary}
\label{sum}

This work presents the exact QME--DBO theory, for which the
time-dependent external field interaction is considered explicitly.
It shows explicitly that
the correlated driving and dissipation result in an effective field correction
that enhances the polarization. This cooperative
property is intrinsically non-Markovian
and reflects the effective bath response to external field applied to the system.
It vanishes in both the Markovian and slow modulation bath limits,
but it is important in the intermediate
bath memory and low-frequency driving region.

For the DBO system, the
correlated effect of driving and non-Markovian dissipation
only appears as a field correction on
the coherent motion.
In a general anharmonic system, this correlated effect
is expected to have much rich implications on the dissipative dynamics
even in the weak system-bath coupling regime, see for example,
Refs.\ \onlinecite{Xu046600} and \onlinecite{Mo05084115}.
The exact non-Markovian dissipative dynamics of general systems under arbitrary
external fields can be studied via the hierarchical
equations of motion approach.\cite{Ish053131,Ish08185,Xu05041103,Xu07031107,Yan04216,Zho08034106}
These include the quantum transport systems, such as the transient current
under the time--dependent bias voltage
reported recently.\cite{Jin08234703,Zhe08093016}

\begin{acknowledgments}
 Support from the National Natural Science Foundation of China (20533060 and 20773114),
 Ministry of Education of China (NCET-05-0546),
 National Basic Research Program of China (2006CB922004),
 and RGC Hong Kong is acknowledged.
\end{acknowledgments}

\appendix

\section{The Wigner phase-space Gaussian wave packet approach}
\label{thWig}

The QME--DBO theory [\Eqs{QMEall} to (\ref{chiep_xi})] can be derived
via the
modified Yan-Mukamel method on the Gaussian wave packet (GWP)
dynamics as follows.\cite{Yan885160,Xu037,Yan05187} The solution to the
Langevin equation of the DBO system reads\cite{Yan05187,Xu037}
\bea\label{QtPt}
        \left[\begin{array}{c} {\hat{q}}(t) \\ {\hat{p}}(t) \end{array}\right]
 &\!\! = &\!\! {\bf T}(t)
        \left[\begin{array}{c} {\hat{q}}(0) \\ {\hat{p}}(0) \end{array}\right]
 \nl  &\!\! \   &\!\!
 + \int^{t}_{0}  \!\d\tau\,  {\bf T}(t-\tau)
        \left[\begin{array}{c} 0 \\ \epsilon(\tau) +F(\tau) \end{array}\right],
\eea
with ${\bf T}(t)$  a $2 \times 2$ matrix:
 \be
   {\bf T}(t)\equiv
   \left[\begin{array}{cc}
      M   \dot\chi(t)  &       \chi(t)       \\
      M^2\ddot\chi(t)  &  M\dot\chi(t)
   \end{array}\right].
 \ee
The Langevin equation in the time-local prescription is then
   \be\label{dothatQtPt}
      \left[\begin{array}{c} {\dot{\hat{q}}}(t) \\ {\dot{\hat{p}}}(t) \end{array}\right] =
      \mbox{\boldmath$\Lambda$}(t)
      \left[\begin{array}{c} \hat{q}(t) \\ \hat{p}(t) \end{array}\right] +
      \left[\begin{array}{c} 0 \\ \epsilon_{\rm eff}(t) + F_{\rm eff}(t) \end{array}\right].
   \ee
Here, $\epsilon_{\rm eff}(t)\equiv\epsilon(t)+\delta\epsilon(t)$ and
\begin{align}
  & F_{\rm eff}(t) \equiv F(t)+\int_0^t\!\d\tau\,\chi_\epsilon(t-\tau;t)F(\tau),
  \label{Feffdef} \\ \label{LamdMatrix}
  & \mbox{\boldmath$\Lambda$}(t)=\dot{\bf T}(t){\bf T}^{-1}(t)
  =\left[\begin{array}{cc} 0 & 1/M
          \\  -M\Omega_t^2 & -\Gamma_t
        \end{array}\right],
\end{align}
with $\Omega_t^2$, $\Gamma_t$, and $\delta\epsilon(t)$ being given in \Eqs{allfuns}.

We can then obtain the equations of motion for all
the cumulants of GWP:\cite{Yan05187,Xu037}
\begin{align}
   & \left[\begin{array}{c}  \dot q_t  \\  \dot p_t
   \end{array}\right]=\mbox{\boldmath$\Lambda$}(t)
   \left[\begin{array}{c}  q_t  \\  p_t
   \end{array}\right] +  \left[\begin{array}{c} 0
   \\ \epsilon_{\rm eff}(t) \end{array}\right],
 \label{dotQtPt} \\ \label{dotWori}
 & \  \dot{\bf W}_t
 = \mbox{\boldmath$\Lambda$}(t){\bf W}_t
        + {\bf W}_t\mbox{\boldmath$\Lambda$}^T(t)
 \nl &  \qquad\ \  +{\rm Re}
    \left[\begin{array}{cc}
      0 &  \la F_{\rm eff}(t)\delta\hat{q}(t)\ra_0
    \\ \la F_{\rm eff}(t)\delta\hat{q}(t)\ra_0
        & 2 \la F_{\rm eff}(t)\delta\hat{p}(t)\ra_0
   \end{array}\right].
\end{align}
Here, $     {\bf W}_t \equiv
 \tiny{\left[\begin{array}{cc}
        \sigma_{qq}(t)   &
        \sigma_{pq}(t)  \\
        \sigma_{pq}(t)   &
        \sigma_{pp}(t)
      \end{array}\right]}$
denotes the phase-space variances of GWP,
$\delta\hat{q}(t) \equiv \hat{q}(t) - q_t$,
$\delta\hat{p}(t) \equiv \hat{p}(t) - p_t$,
and $\la\,\cdot\,\ra_0\equiv{\rm
tr}_{\M}[\,\cdot\,\rho_{\rm T}(0)]$.
We also use the initial factorized state
$\rho_{\rm T}(0)=\rho(0)\rho^{\rm eq}_{\B}$.
From \Eqs{QtPt} and (\ref{Feffdef}), we have
\begin{align}
  \left\la F_{\rm eff}(t)
   \left[\begin{array}{c}\delta\hat{q}(t)\\ \delta\hat{p}(t)\end{array}\right]
  \right\ra_0
&=  \int_0^t\!\d\tau\,
   \left[\begin{array}{c} \chi(t-\tau) \\ M\dot\chi(t-\tau)\end{array}\right]
   \Bigl\{C_{\B}(t-\tau)
\nl&\quad
 + \int_0^t\!\d\tau'\,\chi_\epsilon(t-\tau';t)C_{\B}(\tau'-\tau)\Bigr\}.
\nonumber
\end{align}
Here, $C_{\B}(t)$ is the bath correlation function of \Eq{Ct}.
Comparing with \Eqs{zeta} and (\ref{chixi}),
it is easy to find that
${\rm Re}\la F_{\rm eff}(t) \delta\hat{q}(t) \ra_0 =   \hbar \zeta_t$
and
${\rm Re}\la F_{\rm eff}(t) \delta\hat{p}(t) \ra_0 = M \hbar \zeta'_t$.
Substituting those into \Eq{dotWori}, we obtain
\be\label{dotW}
  \dot{\bf W}_t=\mbox{\boldmath$\Lambda$}(t){\bf W}_t
  + {\bf W}_t\mbox{\boldmath$\Lambda$}^T(t)+
  \hbar \left[\begin{array}{cc}  0 & \zeta_t
           \\  \zeta_t &  2 M \zeta'_t
        \end{array}\right].
\ee

Now using the standard Yan-Mukamel method,\cite{Yan885160}
we arrive at the QME--DBO in the Wigner representation:
 \begin{align}\label{fpw}
    \partial_t\rho_{\rm W}   &  =  \Bigl[
  -\frac{\partial}{\partial q}   \frac{p}{M}
  +M\Omega_t^2\frac{\partial}{\partial p}   q
  -\epsilon_{\rm eff}(t)\frac{\partial}{\partial p}
  \Bigr]\rho_{\rm W}
  \nl    &     \quad
       + \Bigl[    \Gamma_t\frac{\partial}{\partial p}   p
         +M\hbar\zeta'_t  \frac{\partial^2}{\partial p^2}
         +\hbar\zeta_t  \frac{\partial^2}{\partial p\partial q}
         \Bigr]\rho_{\rm W}.
 \end{align}
This is equivalent to \Eq{QMEall}.

\section{The Markovian limit}
\label{app_markov}
  In the Markovian white noise limit,
  $\gamma(t) = 2\Gamma\delta(t)$ or
  $\hat\gamma(\omega) = \Gamma$,
the system response function [\Eq{chiwhar}] reduces to
\be\label{chitmar}
 \chi(t\geq 0)=(M\tilde\Omega)^{-1}\sin(\tilde\Omega t)e^{-\Gamma t/2},
\ee
where $\tilde\Omega \equiv (\Omega^2-\Gamma^2/4)^{1/2}$.
It results in
\be\label{mar_result}
 \Omega_t = \Omega,
\ \ \ \Gamma_t = \Gamma, \ \ \ \chi_\epsilon(\tau;t)=0.
\ee
Thus the effective field correction [\Eq{epsil_efft}]
vanishes in this limit.
It is well known that
$\sigma_{pp}^{\rm eq}\,$
diverges in the Markovian white noise limit without further
the high-temperature assumption.\cite{Gra88115,Wei08}
Same situation applies also to $\zeta_t$
and $\zeta'_t$ [cf.\ \Eq{hpqsig}]. We have therefore  the high-temperature
 limit of $a(t)=Mk_{\rm B}T\gamma(t)/\hbar$, so that $\zeta'_t
= \Gamma k_{\rm B} T/\hbar$ and $\zeta_t = 0$. Thus, \Eq{QMEall} is reduced to
\begin{align} \label{phenom1}
   \dot\rho(t)&=
   -\frac{i}{\hbar} [ H_0 - q\epsilon(t), \rho(t)]
    - \frac{i\Gamma}{2\hbar}[q,\{p,\rho(t)\}]
  \nl &\quad
    - \frac{M\Gamma k_{\rm B}T}{\hbar^2} [q,[q,\rho(t)]].
\end{align}
This is the Caldeira--Leggett equation,\cite{Cal83587} which has
also been extended to anharmonic systems with an arbitrary system-bath
coupling mode $Q$ as\cite{Cao973204,Yan982721}
\bea \label{phenom2}
   \dot\rho(t) &\!\! = &\!\!
   -\frac{i}{\hbar} [ H_0 - \hat\mu\epsilon(t), \rho(t)]
    + \frac{\Gamma}{2\hbar^2} [Q,\{[H_0,Q],\rho(t)\}]
  \nl &\!\! \ &\!\!
    -  \frac{\Gamma k_BT}{\hbar^{2}} [Q,[Q,\rho(t)]].
\eea
Note that the system Hamiltonian
involved here is $H_0$ in the Caldeira--Leggett form.

\section{Equilibrium property}
\label{equi}

In this appendix we will show that $\zeta'_t$ and $\zeta_t$
have the following asymptotic expressions,
 \be\label{hpqsig}
  \hbar\zeta'_{t}\stackrel{t>t_c}{\longrightarrow}\Gamma_t\sigma_{pp}^{\rm eq}/M, \ \
   \hbar\zeta_{t}\stackrel{t>t_c}{\longrightarrow}
   M\Omega_t^2\sigma_{qq}^{\rm eq}-\sigma_{pp}^{\rm eq}/M.
 \ee
Here, $t_c$ is the characteristic time of
the Brownian response function
such that $\chi(t>t_c) \approx 0$.
Thus, \Eq{dotW} behaves asymptotically as
 \be\label{dotWeq}
  \dot{\bf W}_t\stackrel{t>t_c}{\longrightarrow}\mbox{\boldmath$\Lambda$}(t)
  ({\bf W}_t-{\bf W}_{\rm eq})+({\bf W}_t-{\bf W}_{\rm eq})
  \mbox{\boldmath$\Lambda$}^T(t).
 \ee
 This guarantees the correct equilibrium property of QME.
 The involving equilibrium phase-space variances are  $\sigma_{pq}^{\rm eq}=0$
 and \cite{Gra88115,Wei08,Yan05187}
\bsube\label{sigma_eq}
 \begin{align}
\sigma_{qq}^{\rm eq} &= \frac{\hbar}{\pi} \int_{-\infty}^{\infty}\!\!\d\omega
      \frac{{\rm Im} [\hat\chi(\omega)]}{1-e^{-\beta\hbar\omega}}\, ,
 \\ \label{sigmappeq}
\sigma_{pp}^{\rm eq} &= \frac{M^2\hbar}{\pi} \int_{-\infty}^{\infty}\!\!\d\omega
 \frac{\omega^2 {\rm Im} [\hat\chi(\omega)]}{1-e^{-\beta\hbar\omega}}\,  .
 \end{align}
\esube

To prove \Eq{hpqsig}, let us recast
\Eq{zeta} with \Eqs{chiep_xi} as follows.
\bsube\label{hpqdetail}
\begin{align}\label{zetatasymp}
 \zeta_t &= M\left[\dot\chi(t)+\Gamma_t\chi(t)\right]
  \int_0^t\!\!\d\tau\,a(t-\tau)\chi(\tau)
   \nl &\quad
   +M\chi(t)\int_0^t\!\!\d\tau\,a(t-\tau)\dot\chi(\tau)
   \nl &\quad
   -M\int_0^t\!\!\d\tau\int_0^t\!\!\d\tau'\,
   \dot\chi(\tau)a(\tau-\tau')\dot\chi(\tau')
   \nl &\quad
   +M\Omega_t^2\int_0^t\!\!\d\tau\int_0^t\!\!\d\tau'\,
   \chi(\tau)a(\tau-\tau')\chi(\tau'),
\\ \label{zetaptasymp}
 \zeta'_t &=  M\Omega_t^2\chi(t)\int_0^t\!\!\d\tau\,a(t-\tau)\chi(\tau)
   \nl &\quad
   +M\dot\chi(t)\int_0^t\!\!\d\tau\,a(t-\tau)\dot\chi(\tau)
   \nl &\quad
   +M\Gamma_t\int_0^t\!\!\d\tau\int_0^t\!\!\d\tau'\,
   \dot\chi(\tau)a(\tau-\tau')\dot\chi(\tau').
\end{align}
\esube
Note that $a(t)={\rm Re}[C_{\B}(t)]/\hbar$ is an even function and related to the
bath spectral density via the fluctuation--dissipation
theorem:\cite{Wei08,Yan05187}
 \be\label{kappaJ}
   a(t) = \frac{1}{2\pi} \int_{-\infty}^{\infty}\!\!\d\omega\,
   \frac{J(\omega)}{1-e^{-\beta\hbar\omega}}\left(e^{i\omega t}+e^{-i\omega t}\right).
 \ee
Consider now the  $t>t_c$ regime where $\chi(t>t_c) \approx 0$,
the first two terms of both \Eqs{zetatasymp} and (\ref{zetaptasymp})
can be neglected, and every upper limit of the integrations can be set to $\infty$.
Then substituting \Eq{kappaJ} into \Eq{hpqdetail}, performing
Fourier transforms, and using the identity
${\rm Im} [\hat\chi(\omega)] = J(\omega) |\hat\chi(\omega)|^2$ with noticing $J(\omega)=M\omega{\rm Re}[\hat\gamma(\omega)]$,
we finally obtain \Eq{hpqsig}.



\end{document}